\documentclass[onecolumn,tightenlines,nofootinbib,preprintnumbers,amsmath,amssymb]{revtex4}
\usepackage[usenames,dvipsnames]{color}
\usepackage{caption2}
\usepackage{dcolumn}
\usepackage{graphicx}
\usepackage{subfigure}
\usepackage{epstopdf}
\usepackage{bm}
\usepackage{lscape}
\usepackage{slashed}
\renewcommand{\arraystretch}{0.9}
\usepackage{setspace}
\begin{document}
\begin{spacing}{1.5}
\title{Analysis of localized CP asymmetry in $\bar{B}_{s}^{0} \rightarrow \pi ^+\pi ^-\pi^{0}(K^{0})$}
\author{ Yan-Lin Zhao$^{1}$\footnote{E-mail: 2020920966@stu.haut.edu.cn}, Gang L\"{u}$^{1}$\footnote{E-mail: ganglv66@sina.com}, Na-Wang$^{1}$\footnote{E-mail: wangna@haut.edu.cn}, Xin-Heng Guo $^{2}$\footnote{Email: xhguo@bnu.edu.cn}}
\affiliation{\small $^{1}$ Institute of Theoretical Physics, School of Sciences, Henan University of Technology, Zhengzhou 450001, China\\
\small $^{2}$College of Nuclear Science and Technology, Beijing Normal University, Beijing 100875, China\\}
\begin{abstract}
An investigation about localized CP asymmetries for the processes of $\bar{B}_{s}^{0} \rightarrow \pi^{+}\pi^{-}\pi^{0} (K^{0})$ is presented in this paper. The innovation of this paper is that there is a consideration of three-particle $\rho^0(770)$, $\omega(782)$ and $\phi(1020)$ interferences effect. Generally, $\omega(782)$ and $\phi(1020)$ both can decay into $\pi ^+\pi ^-$ pair where can cause extremely small contribution from isospin symmetry breaking. Nevertheless, our analysis shows that $\bar{B}_{s}^{0} \rightarrow \pi ^+\pi ^-\pi^{0}(K^{0})$ decay process can bring differential CP asymmetry about $65\%$ ($36\%$) because of isospin symmetry breaking. To better compared with the data from experimental in the future, we integrate CP asymmetry over the invariant mass and obtain localized CP asymmetry value for the decay $\bar{B}_{s}^{0} \rightarrow \pi ^+\pi ^-\pi^{0} (K^{0})$. We find that there is an evident signal about CP asymmetry at invariant mass value $m(\pi^{+}\pi^{-})$ below the mass of $\rho(770)^{0}$ with the decay $\bar{B}_{s}^{0} \rightarrow \pi ^+\pi^{-}\pi^{0}$.
\end{abstract}
\maketitle

\section{\label{intro}INTRODUCTION}
The sources of CP asymmetry in particle physics have caused a great deal of attention since 1964 \cite{JJ1964}. It is possible for CP asymmetry to occur during the decay of a hadron, during the mixing of neutral hadrons, or during the interference between the two processes \cite{TGe2017}. The strong interaction and electromagnetic interaction all satisfy the CP transformation without deformation while CP symmetry breaks down only in the weak interaction, and it has been believed that the root is because there is a complex phase angle in the Cabibbo-Kobayashi-Maskawa (CKM) matrix \cite{NCa1963}. The amplitude is connected at least twice interact of a weak phase and a strong phase in a b hadron decay \cite{{MDA1979}}. The weak phase is relevant to the Cabibbo-Kobayashi-Maskawa (CKM) matrix while the strong phase is created by rescattering or by other mechanism involved in two-body decay process. The three-body decay process includes complex dynamic mechanism and phase space associated with
resonance and non-resonance contributions. It is the intermediate resonance hadrons linked to Breit-Wigner formalism
that give rise to the strong phases accounting for CP asymmetry in three-body decay processes.

Vector meson dominance model (VMD) predicts that the vacuum polarisation of the photon is entirely made up of vector mesons of  $\rho^{0}(770)$, $\omega(782)$ and $\phi(1020)$ \cite{NB1967}. The photon couples to the neutral vector meson which is dominated by a two-pion state when $e^{+}e^{-}$ decay into the pair of $\pi^{+}\pi^{-}$. The transitions of $\omega(782)$ and $\phi(1020)$ decay to $\pi ^+\pi ^-$ pair which originate in isospin breaking related to the mixings of $\omega(782)-\rho^{0}(770)$ and $\phi(1020)-\rho^{0}(770)$. One can combine the intermediate state with the physical states from isospin states by the unitary matrix for the three hadrons. The dynamics mechanism can be obtained from the interference of  $\rho^{0}(770)$, $\omega(782)$ and $\phi(1020)$ mesons \cite{G2022}. A new strong phase is formed with the help of the intermediate resonance hadrons, which may have an impact on the CP asymmetry of hadron decay.

The LHCb collaboration has gained a lot of attention to the CP asymmetries about $B^{\pm}\rightarrow \pi ^{\pm}\pi ^+\pi ^-$ and $B^{\pm}\rightarrow K^{\pm}\pi ^+\pi ^-$ decay processes in recent years \cite{R.2014,I2020}. Experimental studies have found the above decay processes produce huge CP asymmetry, which reach 58\% and 67\% for the local regions, respectively \cite{RA2013,RA201411}. Currently, there are many theoretical explanations for the CP asymmetry of these three-body decay processes, for example, based on resonance effects and end-state rescattering \cite{ZXY2013,BMJ2013}. Especially for $B^{\pm}\rightarrow K^{\pm}\pi ^+\pi ^-$ process, the huge CP asymmetry occurs when the invariant mass of $\pi^{+}\pi^{-}$ is in region of the $\rho^0(770)$ and $f_0(980)$ resonance \cite{R2014}. Although our previous results show that when the invariant masses of $\pi^{+}\pi^{-}$ in the resonance region of $\rho^0(770)$ and $\omega(782)$ have large CP asymmetry, we failed to compare with the experiment because we do not consider the unevenness of phase space distribution \cite{GLZ2008,LYW2011,LWZ2013,LZX2013,LCG2014}. At the same time, end-state rescattering can produce new strong phase (which can even be relatively large), which may affect CP asymmetry \cite{CCA2005,CCA12005}. Based on the above considerations, we focus on the CP asymmetry in the resonance region of $\rho^0(770)$, $\omega(782)$ and $\phi(1020)$ in this work so that we can make a comparison with the experimental data.

An observation of CP asymmetry has been made in localized phase space region from three-body decay of the B meson \cite{WZWG2015,QWZW2019}. Especially, quasi-two-body $B\rightarrow PV$ resulting in three-body final states from vector meson decays are presented \cite{LLSW2007}. In view of vector mesons resonances, the different types of resonant contributions are allowed to estimate the strong phase. CP violations are measured in charmless B decay which related to the $\rho(770)^{0}-\omega(782) $ mixing region. Including the $\omega(782)$ contribution, the CP violation related to the vector resonance is measureed to be $A_{CP}(B^{+}\rightarrow \rho(770)^{0}\pi^{+}\rightarrow \pi^{+}\pi^{+}\pi^{-})=-0.004 \pm 0.017$ and $A_{CP}(B^{+}\rightarrow\rho^{0}(770)K^{+}\rightarrow K^{+}\pi^{+}\pi^{-})=0.150 \pm 0.019$. The method comes from the the approximation of a two-body interaction plus one spectator meson \cite{LHC2022arxiv}.

These processes improves our knowledge about the CP asymmetry by precise measurements.
CP asymmetry measurements will soon be possible for the $B_{s}$  decay with the data obtained by the upgraded LHCb. The three-body decay of $B_{s}$ meson can provide new opportunities for searching CP asymmetry. In the case of three-body decays, intermediate states often dominate through quasi-two-body decay channels. There are several factorization techniques for calculating the hadron matrix elements in two-body hadron decays \cite{MG1999,MG2000,MM2003,XY2014}.
And by introducing a sudakov factor based on the QCD correction, the PQCD method can calculate decay amplitudes without endpoint divergence is more appreciate. Besises, the final state of this work can be handled as this quasi-two-body form $\bar{B}_{s}^{0} \rightarrow V\pi^{0}(K^{0})$,where $V=\rho^{0}(770),\omega(782),\phi(1020)$ \cite{A1980,A1981,I1981}.

As presented in our work, we consider $\rho^{0}(770)$, $\omega(782)$ and $\phi(1020)$ resonance effect for CP asymmetry, which relates with new complex strong phase. The new strong phase for the first order causes the isospin symmetry to shatter, which results in the CP asymmetry. We aim at the CP asymmetry in the decay process of $\bar{B}_{s}^{0}\rightarrow \rho^{0} \left( \omega ,\phi \right) \pi^{0}(K^{0})\rightarrow \pi ^+\pi ^-\pi^{0}(K^{0})$ in perturbative QCD approach. While at the same time, the localized integrated CP asymmetry can be obtained to compare with the results of experiments in near future.

We organize the information as below. In Section II, theoretical framework about three-particle mixing is introduced. Following that, in Section III, we show the decay diagrams and analytical formalization of the primary decay processes. Section IV of this paper contains the precise calculations, while in Section V we show the numerical findings. Section VI has a summary and a conclusion. Acknowledge is below in Section VII.

\section{\label{sec:cpv1}Three-PARTICLE MIXING}
Using a dominant model of vector meson (VMD), vector mesons of $\rho^0(770)$, $\omega(782)$, $\phi(1020)$ can be obtained by the decay of photons formed by the polarization of positive and negative electron pairs in a vacuum. The momentum of this process is also transmitted through VMD model. Since  the intermediate resonance state is not a physical field, we use the unitary matrix R to make a convert : $\rho _{I}^{0}\left( \omega _I,\phi _I \right)\rightarrow\rho ^0\left( \omega ,\phi \right) $. In the two representations, one can get two expressions about R, where $\left< \rho _I|\rho \right>$, $\left< \omega _I|\omega \right>$ and $\left< \phi _I|\phi \right>$ are equal to 1 and $\left< \rho _I|\omega \right> $, $\left< \rho _I|\phi \right> $ and $\left< \omega|\phi \right> $ are equal to $A_{\rho \omega} \left( s \right)$, $B_{\rho \phi} \left( s \right)$ and $C_{ \omega\phi }\left( s \right)$ since they are all order $\mathcal{O} \left( \lambda \right) ,\left( \lambda \leqslant 1 \right)$ \cite{G2022}. The isospin basis vector $\mid I,I_3>$ can be constructed by using the isospin filed $\rho _{I}^{0}\left( \omega _I,\phi _I \right)$ , where I and $I_3$ refer to the isospin and the isospin third component, respectively. Therefore, we regard physical particle states as linear combinations of these essential vectors. We use the orthogonal normalization method to obtain the relationship between the physical state of the particle and the isospin basis vector, where we define the propagator $D_{\left( s \right)}$. We can get the physical states $\rho^0$, $\omega$ and $\phi$ as follows:
\begin{eqnarray}
\rho ^0=\rho _{I}^{0}-A_{\rho \omega}\left( s \right) \omega _I-B_{\rho \phi}\left( s \right) \phi _I,
\end{eqnarray}
\begin{eqnarray}
\omega =A_{\rho \omega}\left( s \right) \rho _{I}^{0}+\omega _I-C_{\omega \phi}\left( s \right) \phi _I,
\end{eqnarray}
\begin{eqnarray}
\phi =B_{\rho \phi}\left( s \right) \rho _{I}^{0}+C_{\omega \phi}\left( s \right) \omega _I+\phi _I.
\end{eqnarray}

Based on the physical representation, we take the method of diagonalization about $W_I$ by the matrix R without considering higher order terms, where $W_I$ is defined as the mass squared operator in an isospin field. Then we obtain the relationship $A_{\rho \omega} \left( s \right)$, $B_{\rho \phi} \left( s \right)$ and $C_{ \omega\phi }\left( s \right)$ with $W_I$. According to the representations of physics, we can describe the propagator of intermediate state particles from vector mesons. Considering the physics and isospin effect, we make these definitions respectively as follows, where $D_{V_1V_2}=\left< 0|TV_1V_2|0 \right> $ and $D_{V_1V_2}^{I}=\left< 0|TV_{1}^{I}V_{2}^{I}|0 \right>$ (V is vector meson). As a result, we can get $D_{\rho \omega}$, $D_{\rho \phi}$ and $D_{\omega \phi}$, where $D_{\omega \rho}=D_{\rho \omega}$, $D_{\rho \phi}=D_{\phi \rho}$ and $D_{\omega \phi}=D_{\phi \omega}$. Physically, we can ignore all of them since $D_{\rho \omega}$, $D_{\rho \phi}$ and $D_{\omega \phi}$ are equal to zero due to without $\rho-\omega-\phi$ mixing. Besides, the parameters of $\Pi _{\rho \omega}$, $\Pi _{\omega \phi}$, $\Pi _{\rho \phi}$, $A_{\rho \omega}\,\,$, $B_{\rho \phi}\,\,$ and $C_{\omega \phi}\,\,$ are first order approximation. The product which any two of them is higher order and can be disregarded, which enables us to get
\begin{eqnarray}
A_{\rho \omega}=\frac{\Pi _{\rho \omega}}{s_{\rho}-s_{\omega}},~~~
B_{\rho \phi}=\frac{\Pi _{\rho \phi}}{s_{\rho}-s_{\phi}},~~~
C_{\omega \phi}=\frac{\Pi _{\omega \phi}}{s_{\omega}-s_{\phi}}.
\end{eqnarray}
where $s_\rho$, $s_\omega$ and $s_\phi$ refer to the propagators of $\rho$, $\omega$ and $\phi$.

We can write $s_V+m_{V}^{2}-im_V\varGamma _V=s$, where the $\sqrt{s}$ is the invariant mass of two pions. For the vector V, $s_V$ is the inverse propagator, $m_V$ represents its mass and $\varGamma _V$ refers to the decay rate. From above equations, we obtain $\tilde{\Pi}_{\rho \omega}$ and $\tilde{\Pi}_{\rho \omega}$ after definition about $\tilde{\Pi}_{V_1V_2}=\frac{s_{V_1}\Pi _{V_1V_2}}{s_{V_1}-s_{V_2}}$ ($\tilde{\Pi}_{\omega \phi}$ can be ignored in the next step of the calculation).

Wolfe and Maltman recently calculated the precise $\rho -\omega$ mixing parameters \cite{CE2009,CE2011}:
\begin{eqnarray}
\begin{aligned}
	\mathfrak{R} \mathfrak{e} \varPi _{\rho \omega}(m_{\rho}^{2})&=-4470\pm 250_{modl}\pm 160_{data}\mathrm{MeV}^2,\\
	\mathfrak{I} \mathfrak{m} \varPi _{\rho \omega}(m_{\rho}^{2})&=-5800\pm 2000_{modl}\pm 1100_{data}\mathrm{MeV}^2.\\
\end{aligned}
\end{eqnarray}
Mixing parameters $\rho -\phi$ near the $\phi$ meson have been given as follows \cite{MN2000}:
\begin{eqnarray}
F_{\rho \phi}=\left( 0.72\pm 0.18 \right) \times 10^{-3}-i\left( 0.87\pm 0.32 \right) 10^{-3}.
\end{eqnarray}

Different mixing parameters relate to the momentum dependence of $\rho-\omega$ mixing and $\rho-\phi$ mixing, which are $\tilde{\varPi}_{\rho \omega}\left( s \right)$ and $\tilde{\varPi}_{\rho \phi}\left( s \right)$ respectively. To absorb the contribution of $\omega$ and $\phi$ decay into $\pi ^+\pi ^-$  pair, we anticipate discovering evidence of mixing in the resonance area of $\omega$ and $\phi$ where two pions are also generated by breaking isospin symmetry. We use $\tilde{\varPi}_{\rho \omega}\left( s \right) =\Re e\tilde{\varPi}_{\rho \omega}\left( m_{\omega}^{2} \right) +\mathfrak{I} m\tilde{\varPi}_{\rho \omega}\left( m_{\omega}^{2} \right) \,\,$ and $\,\,\tilde{\varPi}_{\rho \phi}\left( s \right) =\Re e\tilde{\varPi}_{\rho \phi}\left( m_{\phi}^{2} \right) +\mathfrak{I} m\tilde{\varPi}_{\rho \phi}\left( m_{\phi}^{2} \right)$ expressions and update the values as below \cite{HB1997,KM1996,SG1998}:
\begin{table}[!ht]
\centering %
\renewcommand{\arraystretch}{2.0} %
\setlength{\tabcolsep}{6.5mm}
\begin{tabular}{|c|c|}
\hline
$\Re e\tilde{\varPi}_{\rho \omega}\left( m_{\omega}^{2} \right) =-4760\pm 440 MeV^2\,\,$ &$\mathfrak{I} m\tilde{\varPi}_{\rho \omega}\left( m_{\omega}^{2} \right) =-6180\pm 3300 MeV^2$
\\ \hline
$\Re e\tilde{\varPi}_{\rho \phi}\left( m_{\omega}^{2} \right) =796\pm 312 MeV^2\,\,$ &$\mathfrak{I} m\tilde{\varPi}_{\rho \phi}\left( m_{\phi}^{2} \right) =-101\pm 67 MeV^2$
\\ \hline
\end{tabular}
\end{table}

\section{\label{sec:cpv1}CP asymmetry in $\bar{B}_{s}^{0}\rightarrow \rho \left( \omega ,\phi \right) \pi^0(K^0)\rightarrow \pi ^+\pi ^-\pi^0(K^0)$}
\begin{figure}[htbp]
       \centering
       \subfigure[\label{fig:a}]{
               \includegraphics[height=3.2cm,width=5cm]{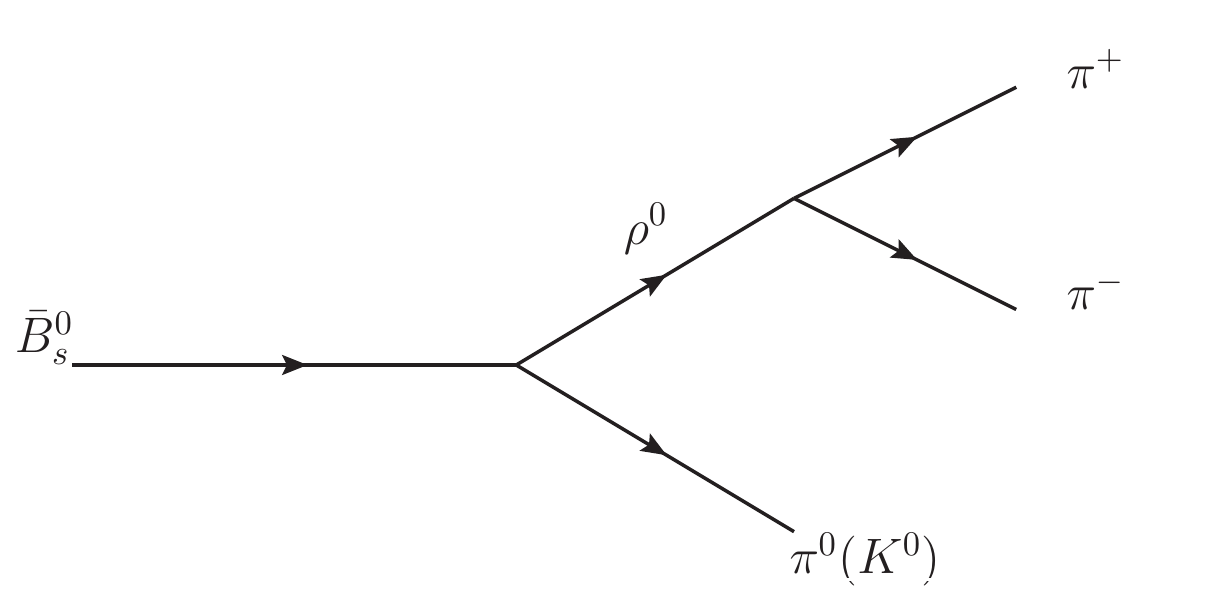}}
       \subfigure[\label{fig:b}]{
               \includegraphics[height=3.2cm,width=5cm]{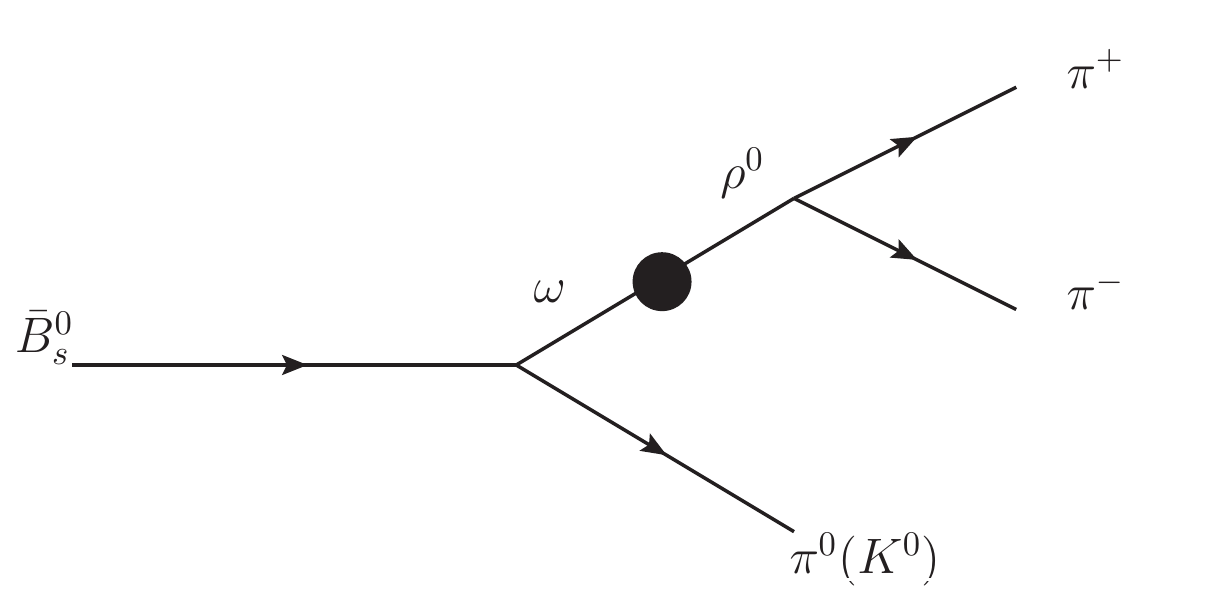}}
       \subfigure[\label{fig:c}]{
               \includegraphics[height=3.2cm,width=5cm]{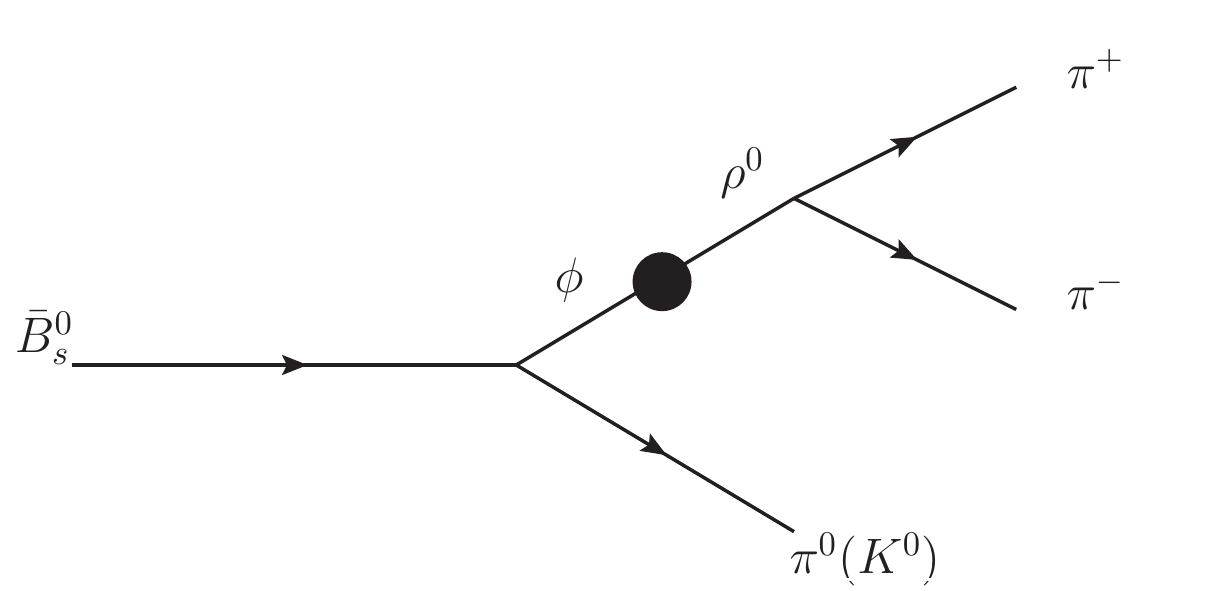}}
       \\
        \subfigure[\label{fig:d}]{
               \includegraphics[height=3.2cm,width=5cm]{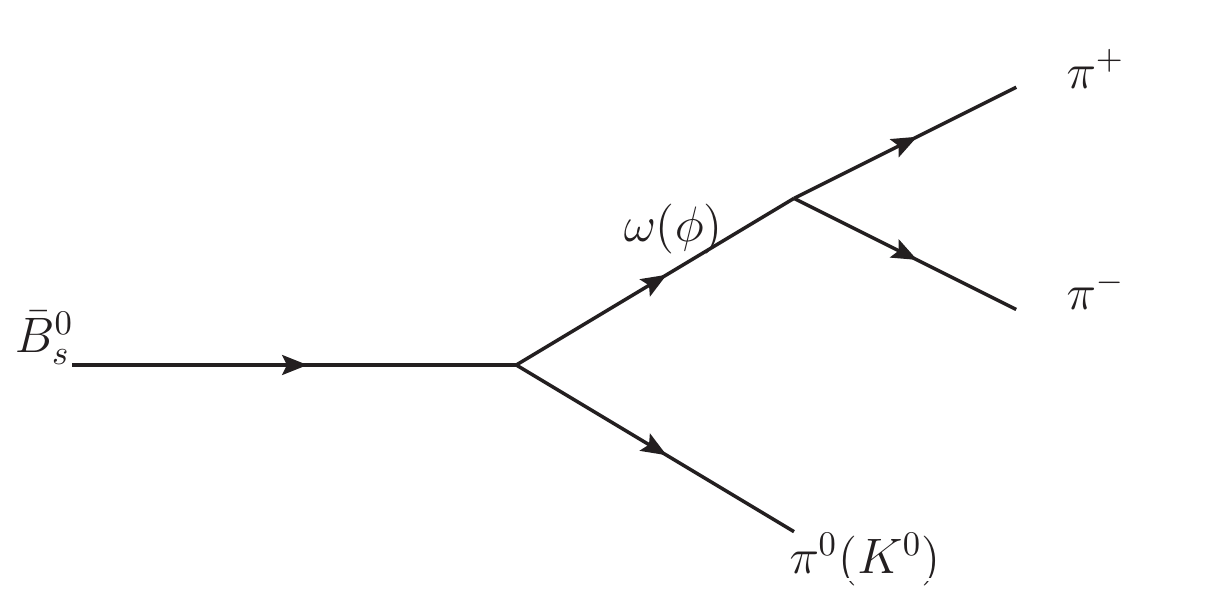}}
        \subfigure[\label{fig:e}]{
               \includegraphics[height=3.2cm,width=5cm]{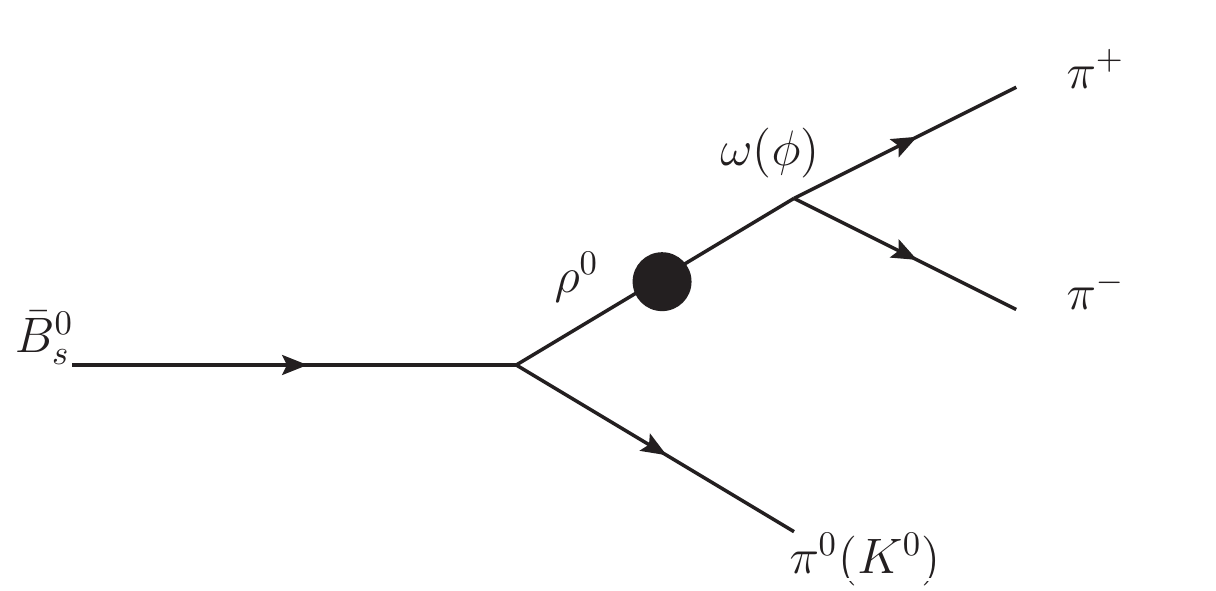}}
        \subfigure[\label{fig:f}]{
               \includegraphics[height=3.2cm,width=5cm]{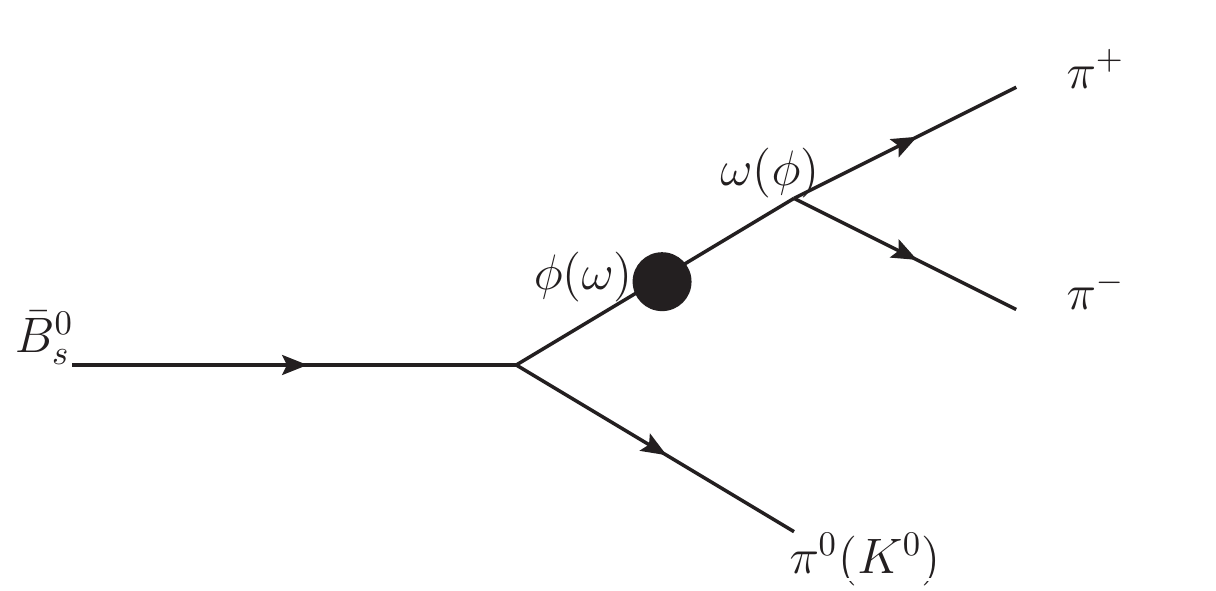}}
        \caption{The diagrams of $\bar{B}_{s}^{0}\rightarrow  \pi ^+\pi ^-\pi^0(K^0)$ decay process.}
        \label{fig2}
        \end{figure}
For the convenience of representing decay process in our work, we replace $\rho^0(770)$, $\omega(782)$, and $\phi(1020)$ with $\rho^0$, $\omega$ and $\phi$ respectively. And the main decay diagrams of $\bar{B}_{s}^{0}\rightarrow \rho \left( \omega ,\phi \right) \pi^0(K^0) \rightarrow \pi ^+\pi ^-\pi^0(K^0)$ can be expressed in Fig.1.  One can see that the quasi-two-body decay of $\bar{B}_{s}^{0}\rightarrow \rho \left( \omega ,\phi \right) \pi^0(K^0)\rightarrow \pi ^+\pi ^-\pi^0(K^0)$ is associated the diagrams $\left( a \right) \thicksim \left( f \right) $ of Fig.1.
The main contribution depends on $(a)$ diagram of the Fig.1 since the decay rate of $\rho \rightarrow \pi \pi $ is $100\%$.
For simplify, we only present a few of the major graphs.

In the diagram (a), $\bar{B}_{s}^{0}$ meson decays into $\pi^0(K^0)$ and $\pi^{+}\pi^{-} $ pair which is produced directly by $\rho^0$ resonance effect. Meanwhile, it is known that $\pi^{+}\pi^{-}$ pair can also exist by the resonance effect of $\omega$ or $\phi$ meson, where corresponding mixing parameters are involved. As shown in the diagram of (b), $\rho^0$ meson decays into $\pi^{+}\pi^{-}$  by $\omega$ resonance. The mixing parameter $\Pi_{\rho\omega}$ is generated during the $\omega$ resonance, which as shown in the black dots of (b). Diagram of (c) is almost similar to diagram (b), but the differences are the resonance effect is $\phi$ and the mixing parameter is $\Pi_{\rho\phi}$. In the diagram (d), these decays of $\omega$ and $\phi$ decay into $\pi^{+}\pi^{-}$ compared with the process of $\rho$ meson is extremely small. Furthermore, the contributions from the diagram (e) and diagram (f) is so tiny that can neglect since $\omega$ and $\phi$ decay into $\pi^{+}\pi^{-}$ through the resonance effect of $\omega-\phi$ minxing. After considering above, we can see $\bar{B}_{s}^{0}\rightarrow \rho \left( \omega ,\phi \right) \pi^0(K^0)\rightarrow \pi ^+\pi ^-\pi^0(K^0)$ decay process receives effectively diagrams contributions from $\left( a \right) \thicksim \left(c \right)$.

The procedure of the decay amplitude $A$($\bar{A}$) is described as:
\begin{eqnarray}
A_{total}=<\pi ^+\pi ^-\pi^0(K^0)|\mathcal{H} |\bar{B}_{s}^{0}>=<\pi ^+\pi ^-\pi^0(K^0)|\mathcal{H} _T|\bar{B}_{s}^{0}>+<\pi ^+\pi ^-\pi^0(K^0)|\mathcal{H} _P|\bar{B}_{s}^{0}>,
\end{eqnarray}
where $<\pi ^+\pi ^-\pi^0(K^0)|\mathcal{H} _T\left( \mathcal{H} _P \right) |\bar{B}_{s}^{0}>|\bar{B}_{s}^{0}>
$ is the contribution of tree (penguin).
Through above contribution, we can define:
\begin{eqnarray}
r\equiv \left| \frac{<\pi ^+\pi ^-\mathrm K^0 |H^P|\bar B_s^{0}>}{<\pi ^+\pi ^-\mathrm K^0 |H^T|\bar B_s^{0}>} \right|.
\end{eqnarray}
The relative value $r$ of the contributions made by the tree operator and the penguin operator are described below:
\begin{eqnarray}
\label{am26}
A=<\pi ^+\pi ^-\pi^0(K^0)|H^T|\bar{B}_{s}^{0}>\left[ 1+re^{i\left( \delta +\phi \right)} \right],
\end{eqnarray}
where $\delta$ and $\phi$ refer to the strong phase and weak phase, respectively.
We can get the physical information from diagrams (a), (b) and (c) in in Fig.1:
\begin{eqnarray}
<\pi ^+\pi ^-\pi^0(K^0)|H^T|\bar{B}_{s}^{0}>=\frac{g_{\rho}}{s_{\rho}s_{\omega}}\overset{\sim}{\Pi}_{\mathrm{\rho\omega}}t_{\omega}+\frac{g_{\rho}}{s_{\rho}s_{\phi}}\overset{\sim}{\Pi}_{\mathrm{\rho\phi}}t_{\phi}+\frac{g_{\rho}}{s_{\rho}}t_{\rho},
\end{eqnarray}
\begin{eqnarray}
<\pi ^+\pi ^-\pi^0(K^0)|H^P|\bar{B}_{s}^{0}>=\frac{g_{\rho}}{s_{\rho}s_{\omega}}\overset{\sim}{\Pi}_{\mathrm{\rho\omega}}p_{\omega}+\frac{g_{\rho}}{s_{\rho}s_{\phi}}\overset{\sim}{\Pi}_{\mathrm{\rho\phi}}t_{\phi}+\frac{g_{\rho}}{s_{\rho}}p_{\rho}.
\end{eqnarray}
where $t_{V}(p_{V})$ is the tree (penguin) amplitudes, $g_{\rho}$ refers to the coupling constant and $s_V$ represents the Breit-Wigner formlism of vector meson($V=\rho,\omega,\phi$).
We define these equations of $\bar{B}_{s}^{0}\rightarrow \rho^0(\omega,\phi) \pi^0(K^0)$ to obtain CP asymmetry according to the Wolfenstein parametrization \cite{LW1964}:\\
\begin{table}[!ht]
\centering %
\renewcommand{\arraystretch}{1.27} %
\setlength{\tabcolsep}{6.5mm}
\begin{tabular}{|c|c|}
\hline
$\bar{B}_{s}^{0}\rightarrow \rho^0(\omega,\phi) \pi^0$ &$\bar{B}_{s}^{0}\rightarrow \rho^0(\omega,\phi) K^0$
\\ \hline
$-\sin \phi \sqrt{\rho ^2+\eta ^2}=\eta$ &$\sin \phi \sqrt{[\rho (1-\rho )-\eta ^2]^2+\eta ^2}=\eta$
\\ \hline
$-\cos \phi \sqrt{\rho ^2+\eta ^2}=\rho$ &$\cos \phi \sqrt{[\rho (1-\rho )-\eta ^2]^2+\eta ^2}+\eta ^2=\rho (1-\rho )$
\\ \hline
\end{tabular}
\end{table}
\\

Therefore, the differential parameter of CP asymmetry is shown as below :
\begin{equation}
\label{cp31}
A_{CP}=\frac{\left| A \right|^2-\left| \overline{A} \right|^2}{\left| A \right|^2+\left| \overline{A} \right|^2}.
\end{equation}

In the future, we need to compare our results with experimental data while taking into consideration the localized integrated direct CP asymmetry of decay processes. For the decay amplitude of $\bar{B}_{s}^{0}\rightarrow \rho^0 \pi^0(K^0)\rightarrow \pi ^+\pi ^-\pi^0(K^0)$ process, we consider the contributions of $\bar{B}_{s}^{0}\rightarrow \rho ^0 \pi^0(K^0)$ and $\rho^0 \rightarrow \pi ^+\pi ^-$.

The amplitude of the $\bar{B}_{s}^{0}\rightarrow \rho^0 \pi^0(K^0)$ can be written as $ M_{\bar{B}_{s}^{0}\rightarrow \rho ^0\pi^0(K^0)}^{\lambda}=\alpha P_{\bar{B}_{s}^{0}}\cdot \epsilon ^*\left( \lambda \right),$ $\alpha$ is the effective coupling constant, $P_{\bar{B}_{s}^{0}}$ is the momenta of
$\bar{B}_{s}^{0}$, $\epsilon $ is the polarization vector of $\rho^0$, $\lambda$ is the helicity of vector meson. And the amplitude for $\rho^0 \rightarrow \pi ^+\pi ^-$ can be written as follows:
$M_{\rho ^0\rightarrow \pi ^+\pi ^-}^{\lambda}=g_{\rho}\epsilon \left( \lambda \right) \cdot \left( p_1-p_2 \right)$, where $g_{\rho}$ is thw effective coupling constant, $\epsilon$ is the plolarization vector of $\rho^0$, $p_1$ is the momenta of $\pi^{+}$ and $p_2$ is the momenta of $\pi^{-}$. Hence, the amplitude of $\bar{B}_{s}^{0}\rightarrow \rho ^0\pi^0(K^0)\rightarrow \pi ^+\pi ^-\pi^0(K^0)$ is \cite{ZH2013,XH2001}:
\begin{eqnarray}
A=\frac{g_{\rho}\alpha}{s_{\rho}}P_{\bar{B}_{s}^{0}}^{\mu}\sum_{\lambda =\pm 1,0}{\epsilon _{\mu}^{*}\left( \lambda \right) \epsilon _r\left( \lambda \right) \cdot \left( p_1-p_2 \right) ^r=-\frac{g_{\rho}\alpha}{s_{\rho}}P_{\bar{B}_{s}^{0}}^{\mu}\left[ g_{\mu r}-\frac{\left( p_1+p_2 \right) _{\mu}\left( p_1+p_2 \right) _r}{m_{\rho}^{2}} \right] \left( p_1-p_2 \right) ^r.}
\end{eqnarray}

In the three body decay process, we obtain $P_{\bar{B}_{s}^{0}}=p_1+p_2+p_3$ and $m_{ij}^{2}=p_{ij}^{2}$ since conservation of energy and momentum. Thus, the amplitude can be written as:
$$
A=\frac{g_{\rho}}{s_{\rho}}\cdot\frac{M_{\bar{B}_{s}^{0}\rightarrow \rho ^0\pi^{0}(K)}^{\lambda}}{P_{\bar{B}_{s}^{0}}.\epsilon ^*}\cdot\left( \xi -s^{'} \right)
\\
=\left( \xi -s^{'} \right) \cdot \mathcal{M},
$$
Using this formula, $\sqrt{s^{'}}$ ($\sqrt{s}$)is the high ( low) invariance mass of the $\pi ^+\pi ^-$ pair. According to the equation above, $\xi$ is equal to $\frac{1}{2}\left( s_{\max}^{'}+s_{\min}^{'} \right)$, where $s_{\max}^{'}$ ( $s_{\min}^{'}$) is the maximum (minimum) value of $s^{'}$ for fixed s.

Perform integral operations on the molecules and denominators of $A_{CP}$ within the invariant mass range, which may be quantified by experiments, we are able to estimate the localized integrated CP asymmetry in this location:
\begin{eqnarray}
A_{CP}^{\varOmega}=\frac{\int_{s_1}^{s_2}{ds\int_{s_{1}^{'}}^{s_{2}^{'}}{ds^{'}}}\left( \xi -s^{'} \right) ^2\left( \left| A \right|^2-\left| \overline{A} \right|^2 \right)}{\int_{s_1}^{s_2}{ds\int_{s_{1}^{'}}^{s_{2}^{'}}{ds^{'}}}\left( \xi -s^{'} \right) ^2\left( \left| A \right|^2+\left| \overline{A} \right|^2 \right)}.
\end{eqnarray}

By analyzing the kinematics of three body decays in the region of $\varOmega \left( s_1<s<s_2,s_{1}^{'}<s<s_{2}^{'} \right) $, we are able to conclude that  $\xi \left[ =\frac{1}{2}\left( s_{\min}^{'}+s_{\max}^{'} \right) \right]$ is correlated with s based on the kinematics. $\xi$ is taken to be a constant in this case because s varies only in minimal scale \cite{R.2014}. This not only results in the cancellation of $\int_{s_{1}^{'}}^{s_{2}^{'}}{ds^{'}}\left( \xi -s^{'} \right) ^2$ but also  $A_{CP}^{\varOmega}$ is independent of the high invariant mass of positive and negative meson pais in this way. We consider the $s$ dependence between the values of $s_{\max}^{'}$ and $s_{\min}^{'}$ in our calculations. It is assumed that $s_{\min}^{'}<s^{'}<s_{\max}^{'}$ represents an integral interval of the high invariance mass of $\pi ^+\pi ^-$, while $\int_{s_{\min}^{'}}^{s_{\max}^{'}}{ds^{'}}\left( \xi -s^{'} \right) ^2$ represents the factor that is dependent upon $s$ \cite{WZWG2015}.

\section{\label{cal}CALCULATION}
In this paper, we take the perturbative QCD (PQCD) method to work out, which is obtained by applying the $k_T$ factorization formalism to the decay process of two-body. Hadrons have a transverse momentum $k_T$, which we must take into account to calculate its transverse momentum. Besides, it introduces an additional energy scale that produces double logarithms in the QCD corrections. As a result, the renormalization group method is a great way to get factors of Sudakov once the terms in this matrix have been resumed. Because of this, the distribution amplitude of mesons in the tiny transverse momentum zone is effectively suppressed in this form factor, which increases the accuracy of the PQCD approach even in this region. The PQCD method has been widely used in the reach of pure annihilation decays. As a result, we believe this approach is appropriate for dealing with processes that are not factorizable and determining the contribution to the annihilation diagram \cite{ying2004}.

In this work, we use the channel of $\bar{B}_{s}^{0}\rightarrow \rho ^0(\omega ,\phi )\pi^0(K^0)\rightarrow \pi ^+\pi ^-\pi^0(K^0)$ as an illustration of how the process conducts under the perturbative QCD. The formalism of $t_{V}$ and $p_{V}$ ($V=\rho,\omega,\phi$), which are obtained from the tree (penguin) level contribution to the equation must be obtained to calculate the CP asymmetry.

Using CKM matrix elements of $V_{ub}V^{*}_{us}$ ($V_{ub}V^{*}_{ud}$) and $V_{tb}V^{*}_{ts}$ ($V_{tb}V^{*}_{td}$) as a basis for analysis, the amplitude in $\bar{B}_{s}^{0}\rightarrow \rho ^0 \pi^0(K^0)\rightarrow \pi^{+}\pi^{-}\pi^0(K^0) $ process can be written as
\begin{eqnarray}
\begin{array}{c}
	A\left( \bar{B}_{s}^{0}\rightarrow \rho ^0\left( \rho ^0\rightarrow \pi ^+\pi ^- \right) \pi ^0 \right) =\frac{G_{F}P_{\bar{B}_{s}^{0}}\cdot \epsilon^*\left( \lambda \right) g_{\rho}\epsilon \left( \lambda \right) \cdot \left( p_{\pi ^+}-p_{\pi ^-} \right)}{\sqrt{2}s_{\rho ^0}}\sum_{\lambda =0,\pm 1}{}\\
	\\
	\begin{array}{c}
	\times \left\{ \,V_{ub}V_{us}^{*} \right. \left[ f_{B_s}F_{ann}^{LL}\left( a_2 \right) +M_{ann}^{LL}\left( c_2 \right) \right] -V_{tb}V_{ts}^{*}\left[ f_{B_s}F_{ann}^{LL} \right.\\
	\\
	\,\,\times \left( a_3+a_9 \right) -f_{B_s}F_{ann}^{LR}\left( a_5+a_7 \right) +M_{ann}^{LL}\left( c_4+c_{10} \right)\\
	\\
	-M_{ann}^{SP}\left( c_6+c_8 \right) -\frac{1}{2}\left. \left. \left( \pi ^+\leftrightarrow \rho ^-+\rho ^+\leftrightarrow \pi ^- \right) \right] \right\} .\\
\end{array}\\
\end{array}
\end{eqnarray}
where the amplitude of $\bar{B}_{s}^{0}\rightarrow \rho ^0 \pi^0$ is composed of two parts $\bar{B}_{s}^{0}\rightarrow \pi ^+\rho ^-$ and $\bar{B}_{s}^{0}\rightarrow \rho ^+\pi ^-$. And they are related to each other as $2A\left( \bar{B}_{s}^{0}\rightarrow \rho ^0\pi ^0 \right) =A\left( \bar{B}_{s}^{0}\rightarrow \pi ^+\rho ^- \right) +A\left( \bar{B}_{s}^{0}\rightarrow \rho ^+\pi ^- \right)$.

\begin{eqnarray}
\begin{array}{c}
	A\left( \bar{B}_{s}^{0}\rightarrow \rho ^0\left( \rho ^0\rightarrow \pi ^+\pi ^- \right) K^0 \right) =\frac{G_FP_{\bar{B}_{s}^{0}}\cdot \epsilon ^*\left( \lambda \right) g_{\rho}\epsilon \left( \lambda \right) \cdot \left( p_{\pi ^+}-p_{\pi ^-} \right)}{2s_{\rho ^0}}\sum_{\lambda =0,\pm 1}{}\\
	\\
	\begin{array}{c}
	\times \left\{ \,V_{ub}V_{ud}^{*} \right. \left[ f_{\rho}F_{B_s\rightarrow K}^{LL}\left( a_2 \right) +M_{B_s\rightarrow K}^{LL}\left( c_2 \right) \right] -V_{tb}V_{td}^{*}\left[ f_{\rho}F_{B_s\rightarrow K}^{LL} \right.\\
	\\
	\,\times \,\left( -a_4+\frac{3}{2}a_7+\frac{1}{2}a_{10}+\frac{3}{2}a_9 \right) +M_{B_s\rightarrow K}^{LR}\left( -c_5+\frac{1}{2}c_7 \right)\\
	\\
	+M_{B_s\rightarrow K}^{LL}\left( -c_3+\frac{1}{2}c_9+\frac{3}{2}c_{10} \right) -M_{B_s\rightarrow K}^{SP}\left( \frac{3}{2}c_8 \right)\\
	\\
	+f_{B_s}F_{ann}^{LL}\left( -a_4+\frac{1}{2}a_{10} \right) -f_{B_s}F_{ann}^{SP}\left( -a_6+\frac{1}{2}a_8 \right)\\
	\\
	+\left. \left. M_{ann}^{LL}\left( -c_3+\frac{1}{2}c_9 \right) +M_{ann}^{LR}\left( -c_5+\frac{1}{2}c_7 \right) \right] \right\} .\\
\end{array}\\
\end{array}
\end{eqnarray}

It is possible to write the decay amplitudes of $\bar{B}_{s}^{0}\rightarrow \omega \pi^0(K^0)\rightarrow\pi^{+}\pi^{-}\pi^0(K^0) $ as follows:

\begin{eqnarray}
\begin{array}{c}
	A\left( \bar{B}_{s}^{0}\rightarrow \omega \left( \omega \rightarrow \pi ^+\pi ^- \right) \pi ^0 \right) =\frac{G_FP_{\bar{B}_{s}^{0}}\cdot \epsilon ^*\left( \lambda \right) g_{\omega}\epsilon \left( \lambda \right) \cdot \left( p_{\pi ^+}-p_{\pi ^-} \right)}{2\sqrt{2}s_{\omega}}\sum_{\lambda =0,\pm 1}{}\\
	\\
	\times \left\{ \,V_{ub}V_{us}^{*} \right. M_{ann}^{LL}\left( c_2 \right) -V_{tb}V_{ts}^{*}\left[ M_{ann}^{LL}\left( \frac{3}{2}c_{10} \right) \right. -M_{ann}^{SP}\left( \frac{3}{2}c_8 \right) +\left. \left. \left( \pi ^0\leftrightarrow \omega \right) \right] \right\} .\\
\end{array}
\end{eqnarray}

\begin{eqnarray}
\begin{array}{c}
	A\left( \bar{B}_{s}^{0}\rightarrow \omega \left( \omega \rightarrow \pi ^+\pi ^- \right) K^0 \right) =\frac{G_FP_{\bar{B}_{s}^{0}}\cdot \epsilon ^*\left( \lambda \right) g_{\omega}\epsilon \left( \lambda \right) \cdot \left( p_{\pi ^+}-p_{\pi ^-} \right)}{2s_{\omega}}\sum_{\lambda =0,\pm 1}{}\\
	\\
	\begin{array}{c}
	\times \left\{ \,V_{ub}V_{ud}^{*} \right. \left[ f_{\omega}F_{B_s\rightarrow K^0}^{LL}\left( a_2 \right) +M_{B_s\rightarrow K}^{LL}\left( c_2 \right) \right] -V_{tb}V_{td}^{*}\left[ f_{\omega}F_{B_s\rightarrow K^0}^{LL} \right.\\
	\\
	\,\,\left( 2a_3+a_4+2a_{5+}\frac{1}{2}a_7+\frac{1}{2}a_9-\frac{1}{2}a_{10} \right) +M_{B_s\rightarrow K^0}^{LR}\left( c_5-\frac{1}{2}c_7 \right)\\
	\\
	+M_{B_s\rightarrow K^0}^{LL}\left( c_3+2c_4-\frac{1}{2}c_9+\frac{1}{2}c_{10} \right) -M_{B_S\rightarrow K^0}^{SP}\left( 2c_6+\frac{1}{2}c_8 \right)\\
	\\
	+f_{B_s}F_{ann}^{LL}\left( a_4-\frac{1}{2}a_{10} \right) +f_{B_s}F_{ann}^{SP}\left( a_6-\frac{1}{2}a_8 \right)\\
	\\
	+\left. \left. M_{ann}^{LL}\left( c_3-\frac{1}{2}c_9 \right) +M_{ann}^{LR}\left( c_5-\frac{1}{2}c_7 \right) \right] \right\}.\\
\end{array}\\
\end{array}
\end{eqnarray}

In order to calculate the amplitude of $\bar B_{s}^{0}\rightarrow \phi \pi^0(K^0)\rightarrow\pi^{+}\pi^{-}\pi^0(K^0)$, we use the following formula:
\begin{eqnarray}
\begin{array}{c}
	A\left( \bar{B}_{s}^{0}\rightarrow \phi \left( \phi \rightarrow \pi ^+\pi ^- \right) \pi ^0 \right) =\frac{G_FP_{\bar{B}_{s}^{0}}\cdot \epsilon ^*\left( \lambda \right) g_{\phi}\epsilon \left( \lambda \right) \cdot \left( p_{\pi ^+}-p_{\pi ^-} \right)}{2s_{\phi}}\sum_{\lambda =0,\pm 1}{}\\
	\\
	\times \left\{ \,V_{ub}V_{us}^{*} \right. \left[ f_{\pi}F_{B_s\rightarrow \phi}^{LL}\left( a_2 \right) +\left. M_{B_s\rightarrow \phi}^{LL}\left( c_2 \right) \right] \right. -V_{tb}V_{ts}^{*}\left[ f_{\pi}F_{B_s\rightarrow \phi}^{LL}\left( \frac{3}{2}a_9-\frac{3}{2}a_7 \right) \right.\\
	\\
	\left. +\left. M_{B_s\rightarrow \phi}^{LL}\left( \frac{3}{2}c_8+\frac{3}{2}c_8 \right) \right] \right\}.\\
\end{array}
\end{eqnarray}

\begin{eqnarray}
\begin{array}{c}
	A\left( \bar{B}_{s}^{0}\rightarrow \phi \left( \phi \rightarrow \pi ^+\pi ^- \right) K^0 \right) =-\frac{G_FP_{\bar{B}_{s}^{0}}\cdot \epsilon ^*\left( \lambda \right) g_{\phi}\epsilon \left( \lambda \right) \cdot \left( p_{\pi ^+}-p_{\pi ^-} \right)}{\sqrt{2}s_{\phi}}\sum_{\lambda =0,\pm 1}{}\\
	\\
	\begin{array}{c}
	\times \left\{ V_{tb}V_{td}^{*} \right. \left[ f_{\phi}F_{B_s\rightarrow K}^{LL}\left( a_3+a_5-\frac{1}{2}a_7-\frac{1}{2}a_9 \right) \right.\\
	\\
	+f_KF_{B_s\rightarrow \phi}^{LL}\left( a_4-\frac{1}{2}a_{10} \right) -f_KF_{B_s\rightarrow \phi}^{SP}\left( a_6-\frac{1}{2}a_8 \right)\\
	\\
	+M_{B_s\rightarrow K}^{LL}\left( c_4-\frac{1}{2}c_{10} \right) +M_{B_s\rightarrow \phi}^{LL}\left( c_3-\frac{1}{2}c_9 \right)\\
	\\
	-M_{B_s\rightarrow K}^{SP}\left( c_6-\frac{1}{2}c_8 \right) -M_{B_s\rightarrow \phi}^{LR}\left( c_5-\frac{1}{2}c_7 \right)\\
	\\
	+f_{B_s}F_{ann}^{LL}\left( a_4-\frac{1}{2}a_{10} \right) -f_{B_s}F_{ann}^{SP}\left( a_6-\frac{1}{2}a_8 \right)\\
	\\
	+M_{ann}^{LL}\left( c_3-\frac{1}{2}c_9 \right) \left. \left. -M_{ann}^{LR}\left( c_5-\frac{1}{2}c_7 \right) \right] \right\} .\\
\end{array}\\
\end{array}
\end{eqnarray}

And $g_V$ is the effective coupling constant of vector meson (V=$\rho$, $\omega$ and $\phi$). $G_F=1.16639\times 10^{-5}GeV^{-2}$ is the Fermi constant, and $a_{i}$ are related to the coefficients of Wilson $c_{i}$. Besides $F_{B_s\rightarrow M_3}$ and $M_{B_s\rightarrow M_3}$ represents the contribution of factorable emission diagrams and annihilation-type diagrams, respectively. LL, LR and SP refer to the contributions from $\left( V-A \right) \otimes \left( V-A \right)$, $\left( V-A \right) \otimes \left( V+A \right)$ and $\left( S-P \right) \otimes \left( S+P \right)$ operators, respectively. There is an non-factorable emission diagram and an annihilation-type diagram in $F_{ann}$ and $M_{ann}$, which indicate their respective contributions. 
The formalisms of the above expressions can be found in the literatures about Perturbative QCD \cite{LLSW2007,ali2007,xiao2011}.

\section{\label{num}Numerical results}
\begin{figure}[!htbp]
\centering
\begin{minipage}[h]{0.45\textwidth}
\centering
\includegraphics[height=4cm,width=6.5cm]{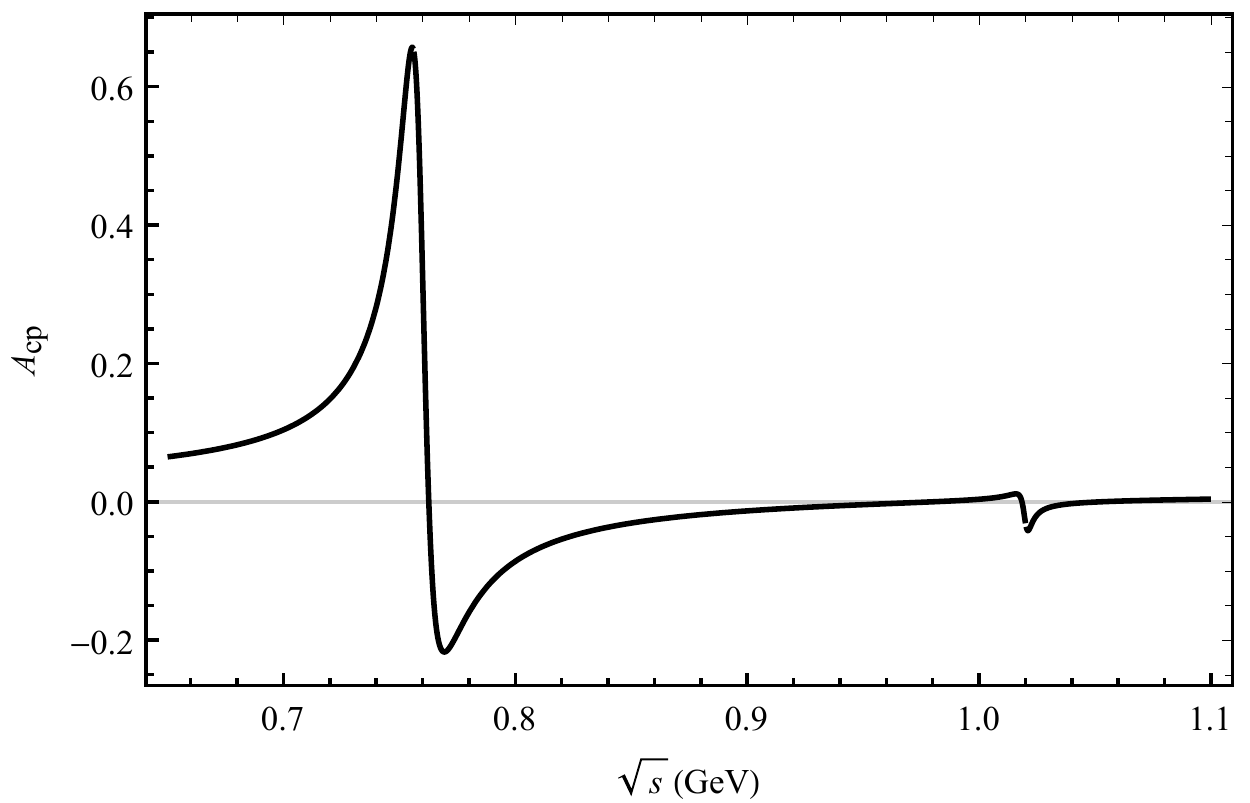}
\caption{Plot of  $A_{CP}$ as a function of $\sqrt{s}$ corresponding to central parameter values of CKM matrix elements
for the decay channel of $\bar{B}_{s}^{0}\rightarrow \pi ^+\pi ^-\pi ^0$.}
\label{fig2}
\end{minipage}
\quad
\begin{minipage}[h]{0.45\textwidth}
\centering
\includegraphics[height=4cm,width=6.5cm]{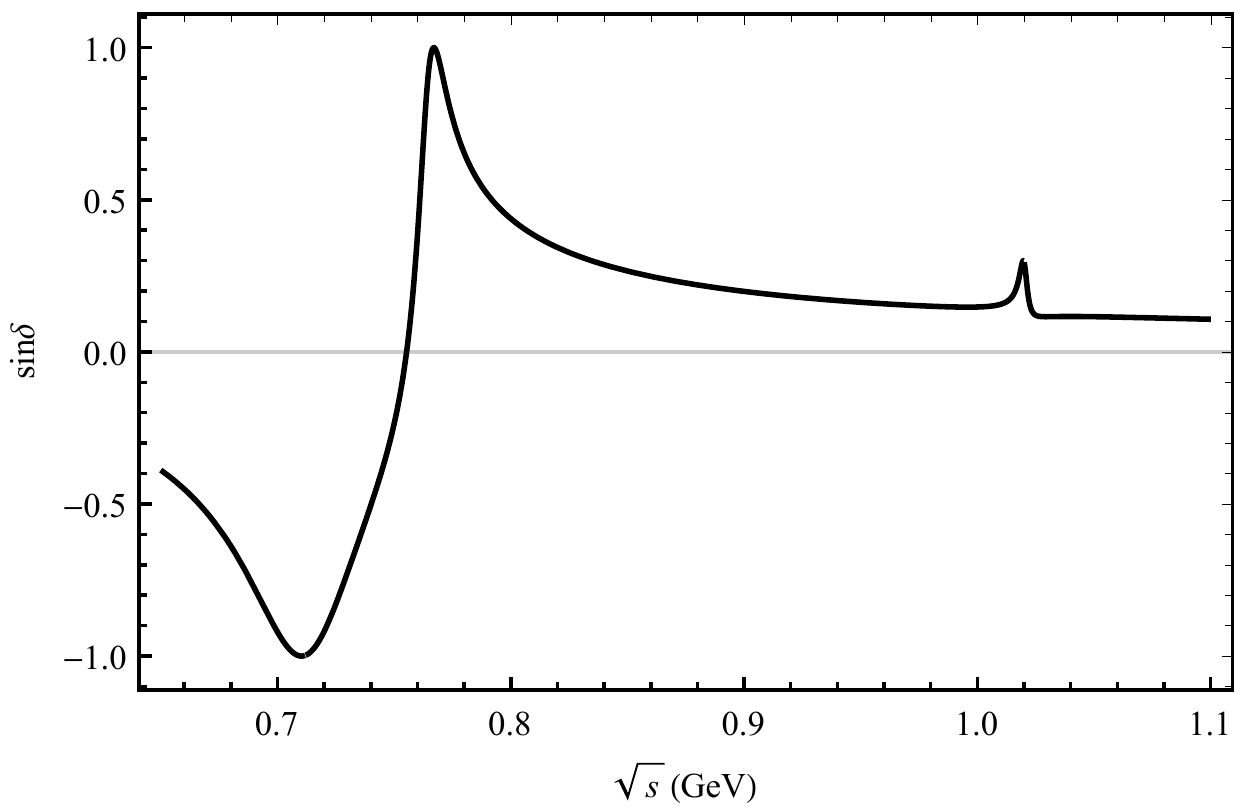}
\caption{Plot of ${\rm{sin}}\delta$ as a function of $\sqrt{s}$ corresponding to central parameter values of CKM matrix elements for the decay channel of $\bar{B}_{s}^{0}\rightarrow \pi ^+\pi ^-\pi ^0$.}
\label{fig3}
\end{minipage}
\end{figure}

\begin{figure}[!htbp]
\centering
\begin{minipage}[h]{0.45\textwidth}
\centering
\includegraphics[height=3.85cm,width=6.25cm]{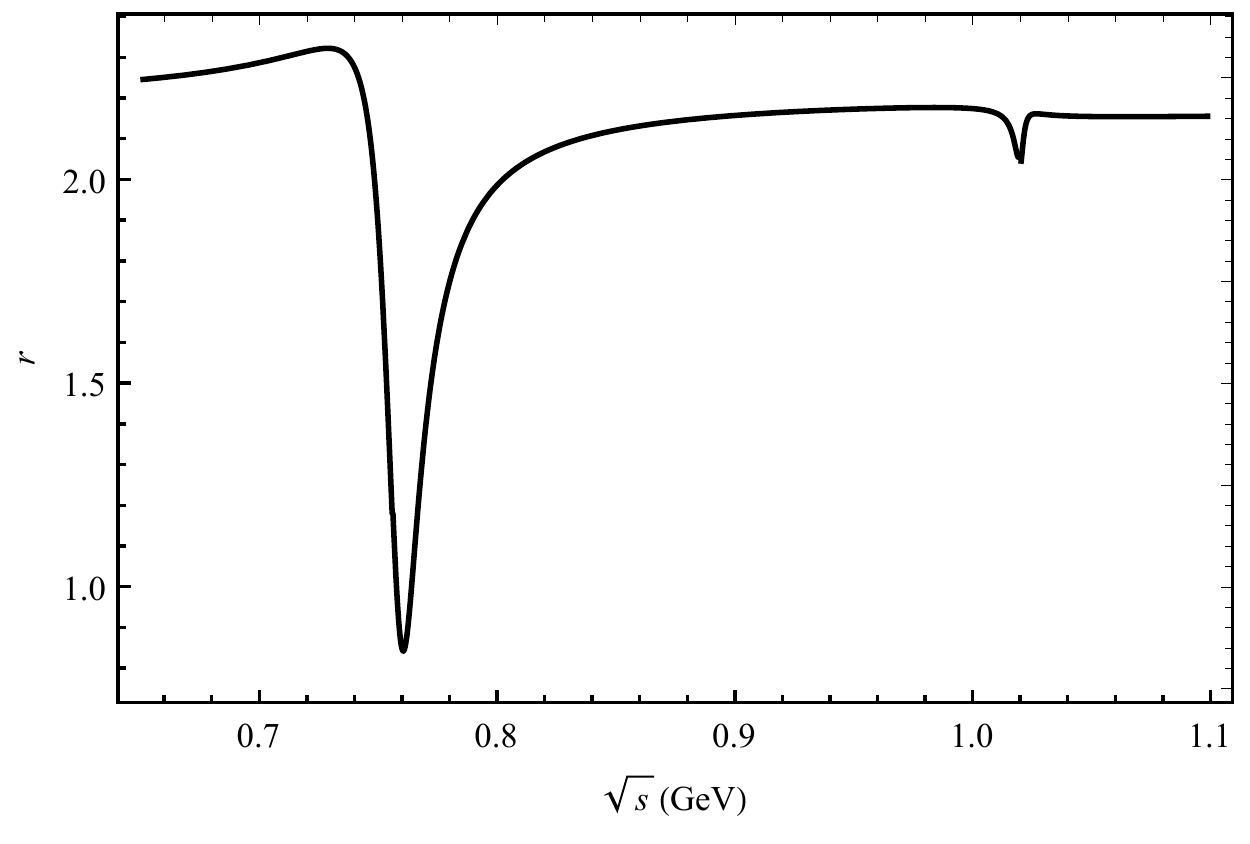}
\caption{Plot of $r$ as a function of $\sqrt{s}$ corresponding to central parameter values of CKM matrix elements for the decay channel of $\bar{B}_{s}^{0}\rightarrow \pi ^+\pi ^-\pi ^0$.}
\label{fig4}
\end{minipage}
\quad
\begin{minipage}[h]{0.45\textwidth}
\centering
\includegraphics[height=4cm,width=6.5cm]{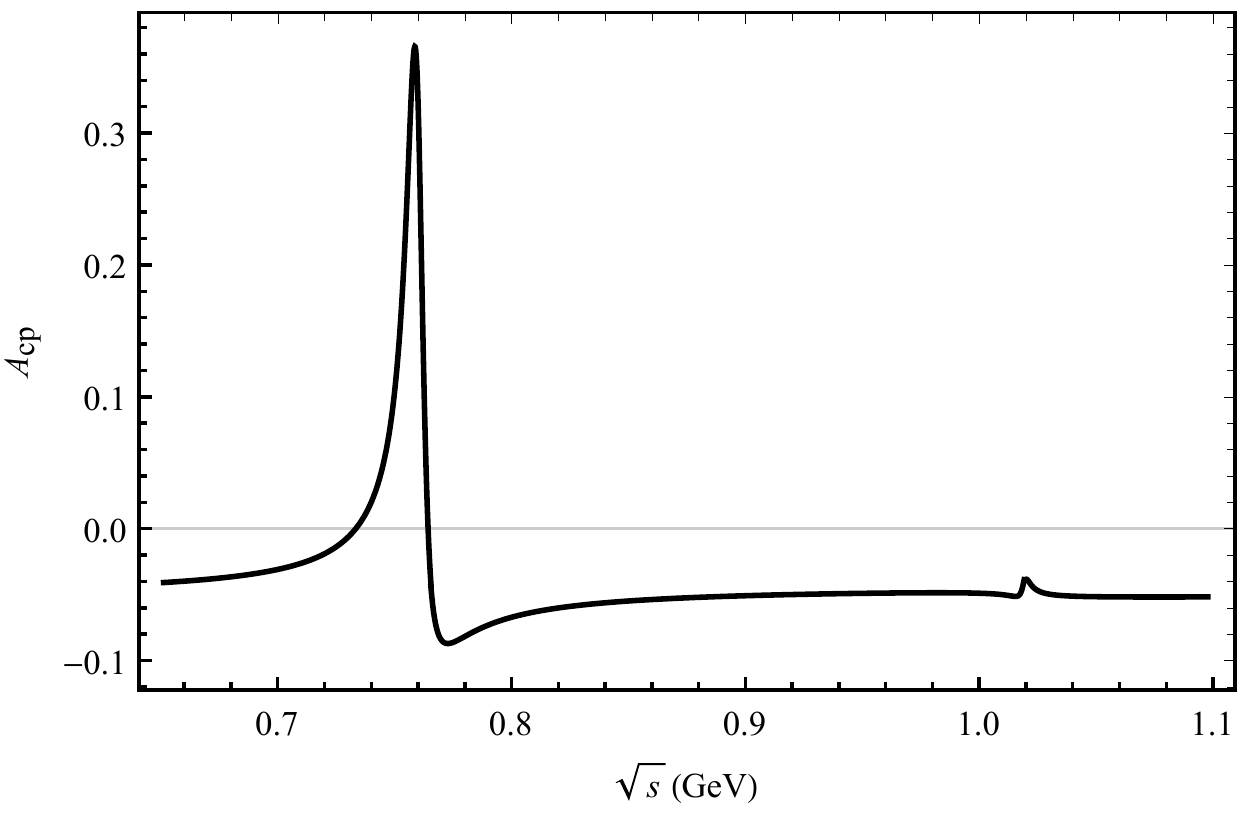}
\caption{Plot of  $A_{CP}$ as a function of $\sqrt{s}$ corresponding to central parameter values of CKM matrix elements for the decay channel of $\bar{B}_{s}^{0}\rightarrow \pi ^+\pi ^-K^0$.}
\label{fig5}
\end{minipage}
\end{figure}

\begin{figure}[!htbp]
\centering
\begin{minipage}[h]{0.45\textwidth}
\centering
\includegraphics[height=4cm,width=6.38cm]{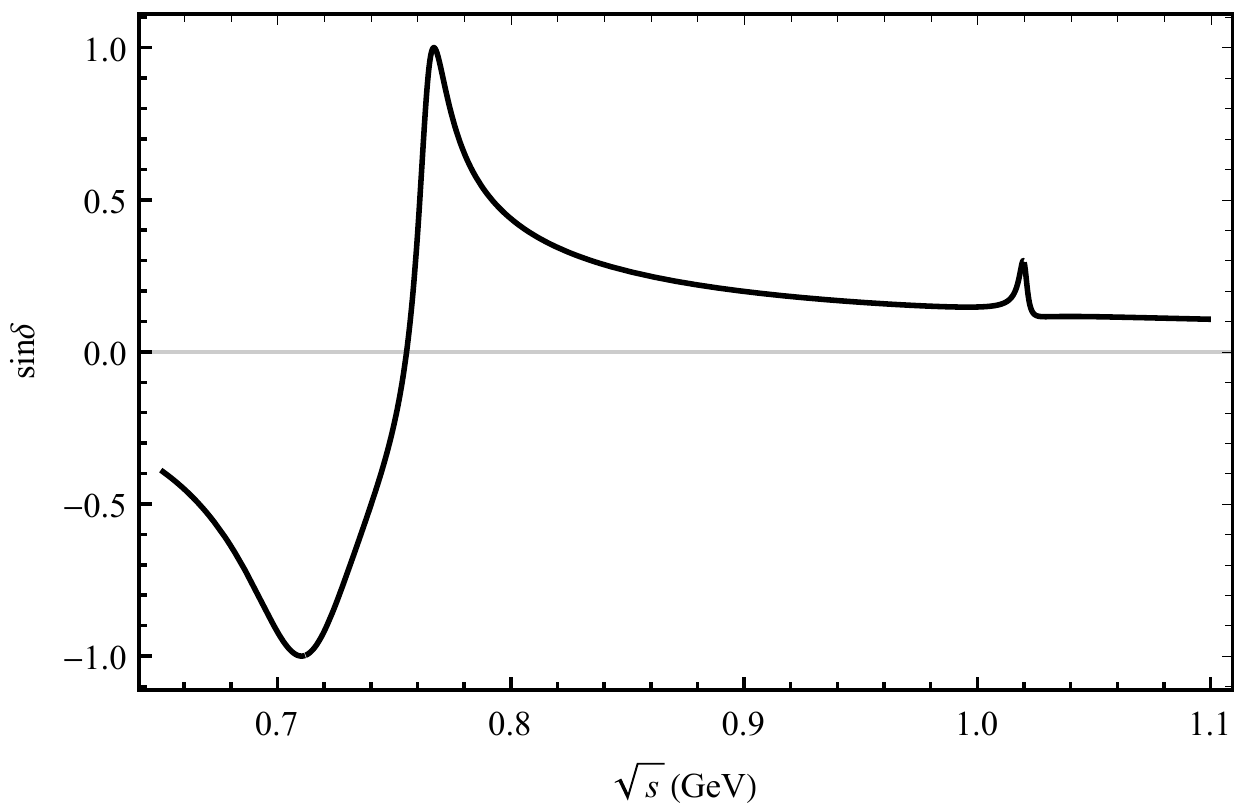}
\caption{Plot of ${\rm{sin}}\delta$ as a function of $\sqrt{s}$ corresponding to central parameter values of CKM matrix elements for the decay channel of $\bar{B}_{s}^{0}\rightarrow \pi ^+\pi ^-K^0$.}
\label{fig6}
\end{minipage}
\quad
\begin{minipage}[h]{0.45\textwidth}
\centering
\includegraphics[height=3.98cm,width=6.1cm]{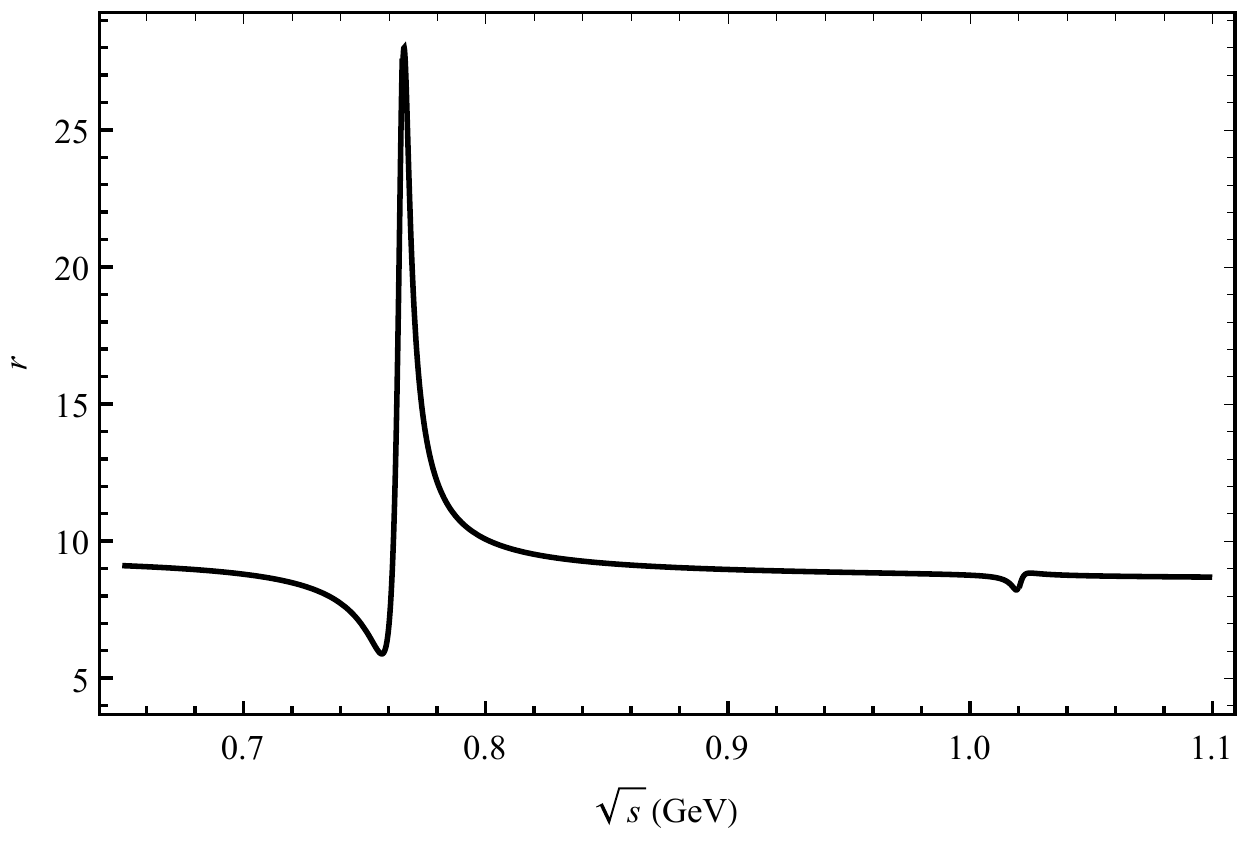}
\caption{Plot of $r$ as a function of $\sqrt{s}$ corresponding to central parameter values of CKM matrix elements for the decay channel of $\bar{B}_{s}^{0}\rightarrow \pi ^+\pi ^-K^0$.}
\label{fig7}
\end{minipage}
\end{figure}

We investigate the CP asymmetry about $\bar{B}_{s}^{0}\rightarrow \pi ^+\pi ^-\pi ^0$ and $\bar{B}_{s}^{0}\rightarrow \pi ^+\pi ^-K^0$ decay processes. According to Eq.(\ref{am26}) and Eq.(\ref{cp31}), we find the CP asymmetry is associated with the weak phase difference, strong phase difference, and $r$. Our results are affected slightly by variation of CKM matrix elements, which determine the weak phase. Thus, the results are presented based on the central parameter values of the CKM matrix elements. For the $\bar{B}_{s}^{0}\rightarrow \pi ^+\pi ^-\pi ^0$ decay process, Fig.2, Fig.3 and Fig.4 show the results. As seen in Fig.2, it is easy to see the CP asymmetry of $\bar{B}_{s}^{0}\rightarrow \pi ^+\pi ^-\pi^0$ channel changes when the invariant masses of the $\pi^+\pi^{-}$ pair is surrounding the $\omega$ resonance and the $\phi$ resonance range, which the maximum CP asymmetry can be achieved $65\%$.

The values of $sin\delta$ and $r$ as a function of $\sqrt{s}$ are shown on the plots of in Fig.3 and Fig.4. If the mass invariant of the $\pi^+\pi^{-}$ pair is in the area where the $\omega$ resonance is located, then one can find that the $sin\delta$ and $r$ vary sharply and see that $sin\delta$ and $r$ vary slightly around the $\phi$ resonance range as compared to the former. With regard to $\bar{B}_{s}^{0}\rightarrow \rho ^0\left( \omega ,\phi \right) \pi ^0\rightarrow \pi ^+\pi ^-\pi ^0$ process, we obtain the CP asymmetries vary from 65\% to -23\% (0\% to -0.05\%) when  invariant mass of the $\pi^+\pi^{-}$ pair is in $\rho -\omega $ ($\rho -\phi $) resonance range.

Fig.5 illustrates the sharp change in CP asymmetry for $\bar{B}_{s}^{0}\rightarrow \pi ^+\pi ^-K ^0$  channel  when the invariant masses of the $\pi^+\pi^{-}$ pairs are around the $\omega$ and $\phi$ resonance range, where the maximum CP asymmetry can be achieved $36\%$. Then, we consider these two situations as follows. When $\pi^+\pi^{-}$ pairs are around region of $\rho-\omega$ mixing, we find the CP asymmetry of $\bar{B}_{s}^{0}\rightarrow \pi ^+\pi ^-K ^0$ can reach 36\%. While the CP asymmetry is just 2\% when $\pi^+\pi^{-}$ pairs is in the mixing of $\rho-\phi$ . Similarly, we analyse the change map about $sin\delta$ and $r$ at the resonance regions in Fig.6 and Fig.7.
Apparent resonance can be found around region of $\rho-\omega$ mixing comparing with slight effect from the $\rho-\phi$ interference.

Concerning the three-body decays, there are complex phase space problems from final states associated with the mechanism of dynamics.
The subsequent $V \rightarrow \pi^{+}\pi^{-}$ is describled as the combination of the coupling constant and the momenta of double $\pi$.
Integration of the phase space provids observation of CP asymmetry for detecting the stucture of intermediate hadrons.
The signals of CP asymmetry are fixed to low mass candidate regions of vectors.
We scan the entire resonance regions to detact the large local CP asymmetry.
Integrating CP asymmetry over the invariant mass of double $\pi$, the localized
CP asymmetry values from -0.006 to -0.01 (-0.014 to -0.02) in $\bar{B}_{s}^{0} \rightarrow \pi ^+\pi ^-\pi^{0}(K^{0})$.
Related to the interferences of the vectors, the desired large CP asymmetry is not emerge,
and cancel each other from the positive and negative values.
The results are in the same order of magnitude as one percent.
However, a clear evident of large CP asymmetry is found  at
invariant mass value $m(\pi^{+}\pi^{-})$ below the mass of $\rho(770)^{0}$ with the decay $\bar{B}_{s}^{0} \rightarrow \pi ^+\pi^{-}\pi^{0}$.
In the regions between  0.65 GeV  and  0.78 GeV, CP asymmetry is consistent to the value of 0.13.
From the 0.78 GeV to 1.1 GeV  region,  CP asymmetry changes sign and reach -0.02.

\section{\label{sum}SUMMARY AND CONCLUSION}
The results of this study illustrate the $\rho-\omega-\phi$ interference caused by the breaking of isospin. Resonance contributions of $\rho-\omega$, $\rho-\phi$ and $\omega-\phi$ can be used to generate a new strong phase. In the processes of $\bar B_{s}^{0}\rightarrow \pi ^+\pi ^-\pi ^0$ and $\bar B_{s}^{0}\rightarrow \pi ^+\pi ^-K^0$, a large CP asymmetry is found to occur in the resonance range. It is possible to reach a maximum CP asymmetry about 65\% in $\bar B_{s}^{0}\rightarrow \pi ^+\pi ^-\pi ^0$ proress. A maximum CP asymmetry of 36\% can occur in $\bar B_{s}^{0}\rightarrow \pi ^+\pi ^-K^0$ channel.

After integration from 0.65 GeV to 1.1 GeV for the decay processes of $\bar B_{s}^{0}\rightarrow \pi ^+\pi ^-\pi ^0$ and $\bar B_{s}^{0}\rightarrow \pi ^+\pi ^-K^0$, we obtain the local CP asymmetry as follows:
\begin{eqnarray}
A_{CP}^{\varOmega}(\bar{B}_{s}^{0}\rightarrow \pi ^+\pi ^-\pi ^0)=-0.008\pm 0.002,
\\
A_{CP}^{\varOmega}(\bar{B}_{s}^{0}\rightarrow \pi ^+\pi ^-K^0)=-0.017\pm 0.003.
\end{eqnarray}

The three-body decay process of bottom and charm mesons is formulated appropriately by the chain decay of quasi-two-body.
We use $B\rightarrow RP_3$ decay process as the case in analyses of quasi-two-body decay. During the progress, R is the state of intermediate resonance state which can further decay to two hadrons $P_{1,2}$, and $P_3$ refers to another final hadron.
This process can be factorized by using the narrow width approximation (NWA), which is also known as the factorization relation. As a result, $B\rightarrow RP_3$ can be written as follows: $\mathcal{B} \left( B\rightarrow RP_3\rightarrow P_1P_2P_3 \right) =\mathcal{B} \left( B\rightarrow RP_3 \right) \mathcal{B} \left( B\rightarrow P_1P2 \right)$ due to the branching ratio. In quasi-two-body decay processes with small widths $\omega$ and $\phi$, the effects can be safely ignored. In light of the large decay rate of $\rho(770)$, it makes sense to carry out a correction. According to the QCD factorization approach, the correction factor for the quasi-two-body decays process of $B^{-}\rightarrow \rho(770)\pi^{-}\rightarrow \pi^{+}\pi^{-}\pi^{-}$ is at level $7\%$. As a measure of the degree of approximation of $\varGamma \left( B\rightarrow RP_3 \right) \mathcal{B} \left( B\rightarrow P_1P_2 \right) =\eta _R\varGamma \left( B\rightarrow RP_3\rightarrow P_1P_2P_3 \right)$, the parameter $\eta _R$ is introduced \cite{chenghaiyang2021prd,chenghaiyang2021plb}. One can ignore the effect of NMA for the calculation of CP violation since the $\eta _R$ can be divided out as a constant. Therefore, we neglect the effects of this correction in this work.

CP asymmetry measurements in the decay of B mesons have become more accurate thanks to the large number of data collected by the LHC in recent years. Theoretical developments using different methods have already led to many predictions of CP asymmetry. The AtLAS and CMS experiments focus on the B physics program and the search of new physics. During the last few years, the LHC has made several upgrades and increased its luminosity by a factor of five.
Based on the amplitude analysis of $B^{+}\rightarrow \pi^{+}\pi^{-}\pi^{-}$ decay, LHCb Collaboration reports different sources of CP asymmetry.
The hadronic structure of intermediate states is sensitive to observation of CP asymmetry.
Contradicting the predictions of the theory, large CP asymmetry connected with the $\rho$  and $\omega$ interference is found, and cancels by intergating the phase spaces. However, there is a evident signal about CP asymmetry at
invariant mass value $m(\pi^{+}\pi^{-})$ below the mass of $\rho(770)^{0}$ from the decay $B^{+}\rightarrow \pi^{+}\pi^{-}\pi^{-}$\cite{R2020prl}.
The search for direct CP asymmetry in charmless $B_{s}$ decay may be measured in the near future.
The interferences $\rho-\omega$, $\rho-\phi$ and $\omega-\phi$ resonances relate to the quasi-two-body decay
$\bar B_{s}^{0}\rightarrow \pi ^+\pi ^-\pi ^0(K^0)$. Analysis of CP asymmetry indicates that the resonance of
$\rho(770)^{0}$ presents the dominant contribution. In fact, the results does not avoide  the interfence of $\rho(770)^{0}$ and  $\omega(782)$
in experiments \cite{LHC2022arxiv}. Hence, the measurement of CP asymmetry includes the effect of $\omega(782)$ meson when one concerns the contribution of $\rho(770)^{0}$ meson. Although, the mass of $\phi(1020)$ is away from the mass of $\rho(770)^{0}$ and  $\omega(782)$ hadrons.
With regard to the SU(3) symmetry of flavour and Isospin symmetry, we can
classify the $\rho(770)^{0}$,  $\omega(782)$ and $\phi(1020)$ hadrons associated with the properties in low mass region of vector interference.
The results manifest that main contribution  is from the resonances of $\rho(770)^{0}$ and $\omega(782)$  in comparison to the interfences of
 $\rho(770)^{0}$- $\phi(1020)$ and $\omega(782)$-$\phi(1020)$ as we expected.

The CP asymmetry can be presented in the regions of $\rho-\omega$ and $\rho-\phi$ regions by reconstructing the $\rho$, $\omega$ and $\phi$ mesons when the invariant masses of $\pi^{+}\pi^{-}$ are at the resonant regions. Hopefully, our predictions will guide future experiments in the right direction.

\section{Acknowledgements}
One of the authors (Gang L) thanks Professor  Zhen-Hua Zhang, Jing-Juan Qi and Chao-Wang for helpful discussions.
This work was supported by National Natural Science
Foundation of China (Project No. 12275024).



\end{spacing}
\end{document}